

\magnification=\magstep1
\centerline{ \bf ON A POSSIBLE SOLUTION FOR THE POLONYI PROBLEM}

\centerline{\bf IN STRING COSMOLOGY}

\vskip 0.5cm
\centerline{M.C. BENTO\footnote{*}{On leave
of absence from Departamento de F\'{\i}sica, Instituto Superior T\'ecnico,
Av. Rovisco Pais, 1096 Lisboa Codex, Portugal.} and O. BERTOLAMI*}
\centerline{Theory Division, CERN,}
\centerline{CH-1211 Geneva 23, SWITZERLAND}
\vskip 0.5cm

\centerline{ABSTRACT}

We establish the main features of homogeneous and isotropic dilaton,
metric and Yang-Mills configurations in a cosmological framework. We identify a
new energy exchange term between the
dilaton and the Yang-Mills field which may lead to a possible solution
of the Polonyi problem in 4-dimensional string models.

\def\ni{\noindent}
\def\laq{\raise 0.4ex\hbox{$<$}\kern -0.8em\lower 0.62 ex\hbox{$\sim$}}
\def\gaq{\raise 0.4ex\hbox{$>$}\kern -0.7em\lower 0.62 ex\hbox{$\sim$}}
\def\cite#1{${}^{#1}$}
\vskip 0.5cm
String theory is the best candidate advanced so far to make General Relativity
compatible with quantum mechanics and  unify all the fundamental
interactions of nature. However, this unification takes place
at very high energy, presumably at the Planck scale, and it is, therefore,
particularly relevant
to study the salient features of this theory in a cosmological
context, hoping to be able to observe some of its implications \cite{1}.
 Four-dimensional string vacua emerging, for instance,
from heterotic string theories, correspond to N=1 non-minimal
supergravity and super Yang-Mills models. The four-dimensional
 low-energy bosonic action arising from string
theory is, at lowest order in $\alpha^\prime$, the string expansion
parameter,  given by

$$
 S_B=\int d^4x \sqrt{-g} \left\{ -{R\over 2 k^2} + 2 (\partial
\phi)^2 - e^{-2 k \phi} Tr\left( F_{\mu\nu} F^{\mu\nu}\right) + 4
V(\phi) \right\} ,\eqno(1)
$$
\ni
where $k^2=8\pi M_P^{-2}$, $M_P$ being the Planck mass and we allow
for a dilaton potential,
$V(\phi)$. The field strength $F_{\mu\nu}^a$ corresponds to the one of
a Yang-Mills theory with a gauge group G, which is a subgroup of $E_8\times
E_8$ or
Spin(32)/$Z_2$. We have set the antisymmetric tensor field
$H_{\mu\nu\lambda}$ to zero and dropped the $F_{\mu\nu}^a \tilde
F^{\mu\nu a}$ term.

As we are interested in a cosmological setting,  we shall focus on  homogeneous
and isotropic
field configurations on a spatially flat spacetime. The most general
metric is then given by

$$
ds^2=-N^2(t) dt^2 + a^2(t) d\Omega^2_3 ,\eqno(2)
$$
\ni
where N(t) and a(t) are respectively the lapse function and the scale
factor.

As for the gauge field, we consider for simplicity the gauge group
G=SO(3);  most of our conclusions, however, will be independent of
this choice. An homogeneous and isotropic ansatz, up to a gauge
transformation, for the gauge potential is the following \cite{2, 3}

$$
 A_\mu (t) dx^\mu = \sum_{a,b,c=1}^{3} {\chi_0(t)\over 4} T_{a b}
\epsilon_{a c b } dx^c,\eqno(3)
$$
\ni
$\chi_0(t) $ being an arbitrary function of time and $T_{ab}$ the generators
of SO(3).

Introducing the ans\"atze (2) and (3) into the action (1) leads, after
integrating over $R^3$ and dividing by the infinite volume of its
orbits, to the  effective action from which our considerations about
the Polonyi problem will follow

$$
S_{eff}=-\int_{t_1}^{t_2} dt \left\{ -{3 {\dot a}^2 a \over k^2 N}
+ {3 a\over N} e^{- 2 k \phi} \left[ {{\dot\chi_0}^2\over 2} - {N^2
\over a^2} {\chi_0^4\over 8}\right] + { 2 a^3\over N} {\dot \phi}^2 - 4
a^3 N V(\phi)\right\},\eqno(4)
$$
\ni
where the dots denote time derivatives.

Let us  consider the so-called entropy crisis and Polonyi problems associated
with the
Einstein-Yang-Mills-Dilaton (EYMD) system. The former  difficulty concerns the
dilution of the baryon
asymmetry generated prior to $\phi$ conversion into radiation.
 The entropy crisis
problem can be solved by  regenerating the baryon asymmetry
after $\phi$ decay, as discussed in  Ref. 4,  considering
models in which the Affleck-Dine mechanism can be implemented to
generate an ${\cal O}(1)$ baryon asymmetry and then allow for its dilution
via $\phi$ decay.

 In
models where the dilaton mass is very small, such that its lifetime is
greater than the age of the Universe ($\Gamma_\phi^{-1}\geq t_U \approx
10^{60}\ M_P^{-1}$), one may  encounter the Polonyi problem, i.e.
$\rho_\phi$ dominates the energy density of the Universe at present.
This problem exists in various N=1
supergravity models with one or more chiral superfields and
even non-minimal models as well as in string models.
A necessary requirement to avoid the problem is that, at the time when $\phi$
becomes non-relativistic, i.e. $H(t_{NR})=m$, the ratio of its energy density
to the one
of radiation satisfies \cite{5}

$$
\epsilon={\rho_\phi(t_{NR})\over
\rho_{\chi_0}(t_{NR})}\laq 10^{-8}.\eqno(5)
$$

Notice that, since the condition  $\Gamma_\phi^{-1}\geq t_U$ implies
$m\leq 10^{-20} M_P$, which falls outside the mass interval for which
inflation takes place (see below), we have to assume that, in models
where this problem occurs,  some  field
other than the dilaton will drive inflation and be responsible for
reheating. Initially and until $\phi$ becomes non-relativistic,
$\rho_\phi\simeq
\rho_{\chi_0} \simeq {1\over 2} m^2 \phi_\ast^2$, implying that
$\epsilon = {\cal O} (1)$ (see e.g. Ref. 5). Hence,  any mechanism for
draining $\phi$ energy into radiation has to be quite effective in
order to be able to help to avoid the  Polonyi problem.
The mechanism we propose is related to the fact that
 our construction does allow
us to describe radiation through the field $\chi_o$ rather than
treating it as a
macroscopic fluid, a fact which has an immediate bearing on the issue
of energy exchange between the dilaton and the Yang-Mills field. In
fact, it is
then easy to show, working out the equations of motion resulting from
the variation of the effective action (4), that

$$
\eqalignno{{\dot \rho}_{\phi} & = - 3 H  {\dot
\phi}^2-  {1\over 2} k e^{- 2 k \phi} \zeta_{\chi_o}
    {\dot \phi}, & (6)\cr
     {\dot \rho}_{\chi_0} & = -4 H \rho_{\chi_0}  +
   6 k {\dot \chi_0^2\over a^2} \dot \phi . &(7)\cr}
$$
\ni
where $H=\dot a / a$ and $ \rho_{\chi_o}  = 3  \left[
  {{\dot \chi_0}^2\over 2 a^2} + {\chi_0^4 \over 8 a^4} \right],
   \zeta_{\chi_0} =  3      \left[
{{\dot \chi}_0^2\over 2 a^2} - {\chi_0^4 \over 8 a^4} \right]$ and $
         \rho_\phi  = { {\dot \phi}^2\over 2} + V(\phi)$.
 The new and somewhat surprising feature  of the above equations  is the
appearance of terms proportional to $\dot \phi$.
The origin of these terms is  ultimately related with the coupling of the
dilaton to the kinetic energy terms of other fields.
Let us then estimate the efficiency of the terms proportional to $\dot \phi$
in Eqs. (6) and (7).  We get for $\epsilon$

$$
\epsilon \simeq {1- 2\Delta /m^2 \phi_\ast^2 \over 1+ 2\Delta^\prime / m^2
\phi_\ast^2},\eqno(8)
$$
\ni
where $\phi_\ast\simeq \phi(t_{RN})\approx M_P$ and  $\Delta$, $\Delta^\prime$
are the integrated contributions of the last two terms
of Eqs. (6) and (7), respectively,  over the time interval ($t_i,\ t_{NR}$).
One expects
that $\Delta\simeq \Delta^\prime$. Demanding $\epsilon$ to
satisfy condition (5) implies that the ratio $\alpha\equiv {2 \Delta
\over m^2 \phi_\ast^2}$ has to be fairly close to 1. From the
equations of motion resulting from the variation of action (4), it is
easy to see
that effective energy exchange occurs
during the period where $H\simeq 2 k \dot \phi$ \cite{1}. Assuming that this
relation holds after inflation and, furthermore, that
$\zeta_{\chi_0}\sim a^{-4}$                       and
$a(t)=  a_R \left({t\over t_R}\right)^{1/2}$, we obtain

$$
\Delta \simeq {t_R^2\over a^4_R}(t_i^{-2} - t_{NR}^{-2}) ,\eqno(9)
$$
\ni
where the index R refers to the time when the inflaton decays.
Hence, in order to get $\alpha={\cal O}(1)$, we must have, if
$t_{NR}\gg t_i$

$$
t_i\simeq {1\over m M_P}{t_R\over a_R^2}.\eqno(10)
$$

 For typical values of the relevant parameters, e.g. $t_i\simeq 10^{10}
M_P^{-1}$, $t_R\gaq 10^{30} M_P^{-1}$ and $ a_R\gaq 10^{30}
M_P^{-1}$, we see that the dilaton mass is required to be exceedingly
small, \break $m\leq 10^{-40} M_P$. Since solving the
Polonyi problem requires $\alpha$ to be very close to 1, it is clear,
from our estimate, that this can be achieved provided the energy
exchange is effectively maintained over a sufficiently long time.
 Actually,   energy exchange via terms
proportional to $\dot \phi$  occurs also when coupling the dilaton to
bosons and fermions through $e^{- 2
k \phi} {\cal L}_{matter}$ (see e.g. Ref. 6), which will then contribute to
further draining of $\phi$ energy. Other contributions to this process
would occur if we had chosen a larger gauge
group as, besides $\chi_0(t)$, another multiplet of fields would
appear in the effective action \cite{2, 3} leading to extra   energy
exchange terms.

Notice that, when discussing a very light or massless dilaton, one has
to deal with the implications of the fact that coupling constants and
masses are dilaton dependent and the ensued problems, such as
cosmological variation of the fine structure constant as well as other
coupling constants and violations of the weak equivalence principle.
The study of the cosmological evolution of the Einstein-Matter-Dilaton
system indicates that the inclusion of
non-perturbative string loop effects is crucial to render
consistency with the experimental data if the dilaton is massless \cite{7}. The
impact of the string loop
effects in the EYMD system is to change the Yang-Mills-Dilaton coupling to
$B(\phi)
F_{\mu\nu}^a F^{\mu\nu a}$, where $B(\phi)=e^{-2k\phi} + c_0 + c_1
e^{2k\phi}  + ...$ and $c_0,\ c_1,...$ are
constants. Hence, in what concerns the Polonyi problem we have been
discussing, the extra  terms
may either weaken or reinforce our previous conclusions regarding a possible
solution to this problem, depending on the value and more crucially
the  sign of the constants $c_i$.

Finally, we briefly describe our results concerning the  inflationary solutions
of the model. As shown in Ref. 6,
where  radiation is treated as a fluid, one obtains chaotic inflationary
solutions driven by the dilaton for   $ V(\phi)= {1\over 2} m^2 (\phi -
\phi_0)^2$, with $ 10^{-8}M_P < m < 10^{-6}M_P$ and $\phi_0\sim M_P$, a
result which
remains valid if we add a quartic term to the potential. Although
the dilaton potential, which has its origin is non-perturbative
effects such as gaugino condensation and a possible v.e.v. for the
antisymmetric tensor field, has a more complicated structure, it
is reassuring to see that it is possible to obtain inflationary
 solutions in
simple cases. We have checked that, with our field treatment of radiation,
inflationary solutions do exist and inflation  with more than 65
e-foldings requires that the initial value of the $\phi$ field is such
that $\phi_{i}\gaq 4 M_P$ \cite{1,3,6} and, actually,  these correspond
  to most of the trajectories, with a probability $1-(m/M_P)^2$.
Inflation is therefore a fairly general feature of expanding  models with
$V(\phi)= {1\over 2}m^2 (\phi-\phi_0)^2$, for $\phi-\phi_0>0$ and  where the
initial value of
$\phi$ satisfies the condition $\phi_{i}\gaq 4 M_P$. In fact, this
initial condition
is indeed shown to be  favoured, as follows from the
study of the solutions of the Wheeler-DeWitt equation for the EYMD system in
the minisuperspace approximation\cite{8}.

\bigskip
\centerline{ References}
\medskip

\item{1.}M.C. Bento and O. Bertolami, ``General Cosmological Features
of the Einstein-Yang-Mills-Dilaton System in String Theories'',  to appear in
{\it Phys. Lett.
B}.

\item{2.} O. Bertolami, J. M. Mour\~ao, R.F. Picken  and I. P.
Volobujev, {\it Int. J. Mod. Phys.\/} {\bf A6 }(1991) 4149.

\item{3.} P.V. Moniz, J.M. Mour\~ao and P.M. S\'a, {\it Class. Quantum
Gravity\/} {\bf 10} (1993) 517.

\item{4.} M.C. Bento, O. Bertolami and P.M. S\'a, {\it Mod. Phys. Lett.\/}
{\bf A7} (1992) 911.

\item{5.} O. Bertolami, {\it Phys. Lett.\/} {\bf B209} (1988) 277.

\item{6.} M.C. Bento, O. Bertolami and P.M. S\'a, {\it Phys. Lett.\/}
{\bf B262} (1991) 11.

\item{7.} T. Damour and A. M. Polyakov, {\it Nucl. Phys.}  {\bf
B423} (1994) 532.

\item{8.} M.C. Bento and O. Bertolami, in preparation.
\bye